\documentclass[preprint]{aastex}
\usepackage{color}

\begin{document}

\title{Discovery of a Two-Armed Spiral Structure in the Gapped Disk in HD 100453}
\color{black}
\shorttitle{HD 100453}
\shortauthors{placeholder}

\author{Kevin Wagner\altaffilmark{1}}
\affil{Department of Astronomy/Steward Observatory, The University of Arizona, 933 North Cherry Avenue, Tucson, AZ 85721}
\altaffiltext{1}{National Science Foundation Graduate Research Fellow}

\email{kwagner@as.arizona.edu}

\author{Daniel Apai\altaffilmark{2}$^{,}$\altaffilmark{3}}

\affil{Department of Astronomy/Steward Observatory, The University of Arizona, 933 North Cherry Avenue, Tucson, AZ 85721}

\author{Markus Kasper}
\affil{European Southern Observatory, Karl-–Schwarzschild-–Str. 2, D–85748 Garching, Germany}
\and
\author{Massimo Robberto}
\affil{Space Telescope Science Institute, 3700 San Martin Dr., Baltimore, MD 21218, USA}

\altaffiltext{2}{Lunar and Planetary Laboratory, The University of Arizona, 1640 E. University Blvd., Tucson, AZ 85718, USA}
\altaffiltext{3}{Earths in Other Solar Systems Team, NASA Nexus for Exoplanet System Science}

\begin{abstract}

We present VLT/SPHERE adaptive optics imaging in Y$-$, J$-$, H$-$, and K-bands of the HD 100453 system and the discovery of a two-armed spiral structure in a disk extending to 0$\farcs$37 ($\sim$42 AU) from the star, with highly symmetric arms to the Northeast and Southwest. Inside of the spiral arms, we resolve a ring of emission from 0$\farcs$18-0$\farcs$25 ($\sim$21-29 AU). By assuming that the ring is intrinsically circular we estimate an inclination of $\sim$34$^{o}$ from face-on. We detect dark crescents on opposite sides (NW and SE) which begin at 0$\farcs$18 and continue to radii smaller than our inner working angle of 0$\farcs$15, which we interpret as the signature of a gap at $\lesssim$21 AU that has likely been cleared by forming planets. We also detect the $\sim$120 AU companion HD 100453 B, and by comparing our data to 2003 HST/ACS and VLT/NACO images we estimate an orbital period of $\sim$850 yr. We discuss what implications the discovery of the spiral arms and finer structures of the disk may have on our understanding of the possible planetary system in HD 100453, and how the morphology of this disk compares to other related objects.

\end{abstract}

\keywords{Stars: individual (HD 100453) -- stars: variables (Herbig Ae/Be) -- planetary systems: protoplanetary disks -- planetary systems: planet-disk interactions}

\section{Introduction}

HD 100453 is an isolated Herbig Ae star in the Lower Centaurus-Crux association (\citealt{malfait98} and \citealt{Kouwenhoven2005}). Hipparcos parallax measurements suggest a distance of 114$^{+11}_{-4}$ pc and proper motion of $\alpha$=$-$36.95$\pm$0.59 mas/yr and $\delta$=$-$4.72$\pm$0.74 mas/yr \citep{perryman97}. The star is of spectral type A9V with L$\sim$9 L$_{\sun}$ and M$\sim$1.7 M$_{\sun}$ \citep{dominik03}. A circumstellar disk is inferred from the infrared spectral energy distribution (SED) of the object (\citealt{meeus01} and 2002). From the far-UV continuum, \cite{collins09} place an upper limit on the mass accretion rate of $\dot{M} \lesssim 1.4\times 10^{-9}$ M$_{\sun}$/yr. The SED shows excess of near infrared and mid infrared emission and can be fit with a power-law plus a blackbody, placing the object among the Meeus group I sources. These objects are inferred to have optically thin flared outer disks surrounding an optically thick disk at the mid-plane, and many have gaps in their structure that are interpreted as signatures of giant planet formation. Most of these gaps are inferred through the SED and have not been directly imaged, raising the possibility that some of these inferred gaps may rather be due to a lower density of small dust grains. The verification of gaps with high contrast imagery at multiple wavelengths is necessary to determine which systems are likely to host recent or on-going planet formation. The SED of HD 100453 also shows the presence of poly-cyclic aromatic hydrocarbon (PAH) emission and cold crystalline silicates \citep{vandenbussche04}, though the spectrum lacks a strong 10 $\mu$m silicate emission feature \citep{meeus01}, which could be further evidence of a gap \citep{maaskant14}. 

Aside from spatially resolved PAH emission \citep{habart06} and Q-band imaging \citep{marinas11}, the disk has never been resolved in high-contrast imagery. VLT/NACO observations in 2003 revealed a M4.0V-M4.5V companion candidate of $\Delta$K$_{s}$=5.1 mag contrast to the host star \citep{chen06}. The companion has been confirmed to be co-moving \citep{collins09} with an age of 10$\pm$2 Myr, consistent with ages found by placing the A-star on PMS evolutionary tracks \citep{meeus02} of t$\gtrsim$10 Myr, and with the age of the stellar association in Lower Centaurus-Crux \citep{mamajek13}, t$\sim$10 Myr. Another faint object detected in the field by both of the aforementioned authors turns out to be a background star as its relative motion vector is opposite in direction to the proper motion of the HD 100453 system. We adopt these authors' convention and refer to this background object henceforth as `star C'.

Several other disks around young stars have been resolved to have spiral structures (e.g. SAO 206462, \citealt{muto12}; MWC 758, \citealt{grady13}, \citealt{benisty15}; Oph IRS 48, \citealt{follette14}), with one interesting interpretation of their formation invoking the gravitational influence of unseen planets. Two spiral arms seem to be the most common (e.g. SAO 206462 and MWC 758), although objects with a single spiral arm have also been detected (Oph IRS 48) or suggested (V 1247 Ori, \citealt{kraus13}). Theoretical studies \citep{pohl15} have shown that wide open spiral arms can be generated as scale height variations due to the gravitational instability induced on the outer disk by a planet of planet-to-star mass ratio of $10^{-2}$, and that spirals induced via this mechanism should be well above the detection limits of current telescopes. The same authors are only able to reproduce two-armed spirals with two planets, and highly symmetric spirals require very specific ratios in mass, radial location, and azimuthal angles of the planets.

In this letter we present the detection of the circumstellar disk in HD 100453 in scattered Y$-$ K-band light using VLT/SPHERE. We describe our observations in $\S$2 and present the results in $\S$3. In $\S$4, we discuss the morphology of the HD 100453 system, and place the spiral structure in the context of the few other disks with double spirals. We also discuss further constraints that our observations place on the nature of other objects in the field (HD 100453 B, star C). We summarize our results in $\S$5.

\section{Observations and Data Reduction}

We observed HD 100453 on 2015 April 10 using VLT/SPHERE at ESO's Paranal Observatory under ESO program 095.C-0389 (PI: D. Apai). Our observations were carried out in IRDIFS extended mode \citep{beuzit08}, allowing for simultaneous use of the infrared dual-band imager and spectrograph (IRDIS; \citealt{dohlen08}) and integral field spectrograph (IFS; \citealt{claudi08}). We used IRDIS to take dual-band images in K1 and K2 filters, centered at 2102 nm and 2255 nm, with $\Delta \lambda$=102 nm and $\Delta \lambda$=109 nm, respectively. Simultaneously, the IFS was used to obtain low-resolution spectra from Y$-$ to H-band with R$\sim$30 between 0.95$-$1.65 $\mu$m. The observations were corrected for atmospheric turbulence and internal wave-front errors using the Extreme Adaptive Optics SAXO system \citep{fusco14}. The signal from the target star was suppressed using an apodized Lyot coronagraph with a mask radius of 0$\farcs$0925 (\citealt{carbillet11} and \citealt{guerri11}), optimized for observations from Y$-$ to H-band with an inner working angle of 0$\farcs$15. Additionally, we collected i) coronagraphic images with four centrally symmetric satellite speckles (created by generating sine aberrations with the deformable mirror) to determine the position of the star behind the mask, and ii) images with the star shifted away from behind the mask to be used for flux calibrations (with a neutral density filter to avoid saturation).\footnote{Our flux calibration measurements included the ND1 filter to avoid PSF saturation. The transmission curves can be found on the SPHERE web-page (http://www.eso.org/sci/facilities/paranal/instruments/sphere.html).} This is standard observing procedure with SPHERE and allows for photometry and astrometry to be calibrated with high fidelity. The plate scales for IRDIS and IFS are 12.251$\pm$0.005 mas/pixel, and 7.46$\pm$ 0.02 mas/pixel, respectively, as provided in the SPHERE user manual.

We collected data for a total of 1408 seconds of integration time with IRDIS and 1600 seconds with IFS. The data were calibrated and stored in master data cubes using the SPHERE pipeline reduction software \citep{pavlov08}, which creates and subtracts the master dark frame, replaces bad pixels (by replacing each bad pixel with the median of its neighbors), and finally divides by the flat field. The data were then corrected for  0.6\% \color{black} anamorphic distortion (Maire et al., submitted). \color{black} Similarly for the IFS, we used the pipeline to create the x-y-$\lambda$ data cubes. Parts of these cubes were stacked to create broad-band images in Y-band (990-1100 nm), J-band (1140-1350 nm) and H-band (1490-1639 nm). A filter cuts off the long-end of the H-band in order to limit sky background. We compared the IRDIS and IFS images to a library of PSFs (from other A stars in our target pool), scaled the PSFs, and subtracted the PSF that minimizes the residuals. The IRDIS K1 and K2 images are presented in Figure 1, whereas wavelength stacked Y$-$, J$-$, and H-band images from the IFS are presented in Figure 2.

\begin{figure}[htpb]
\figurenum{1}
\epsscale{1.2}
\plottwo{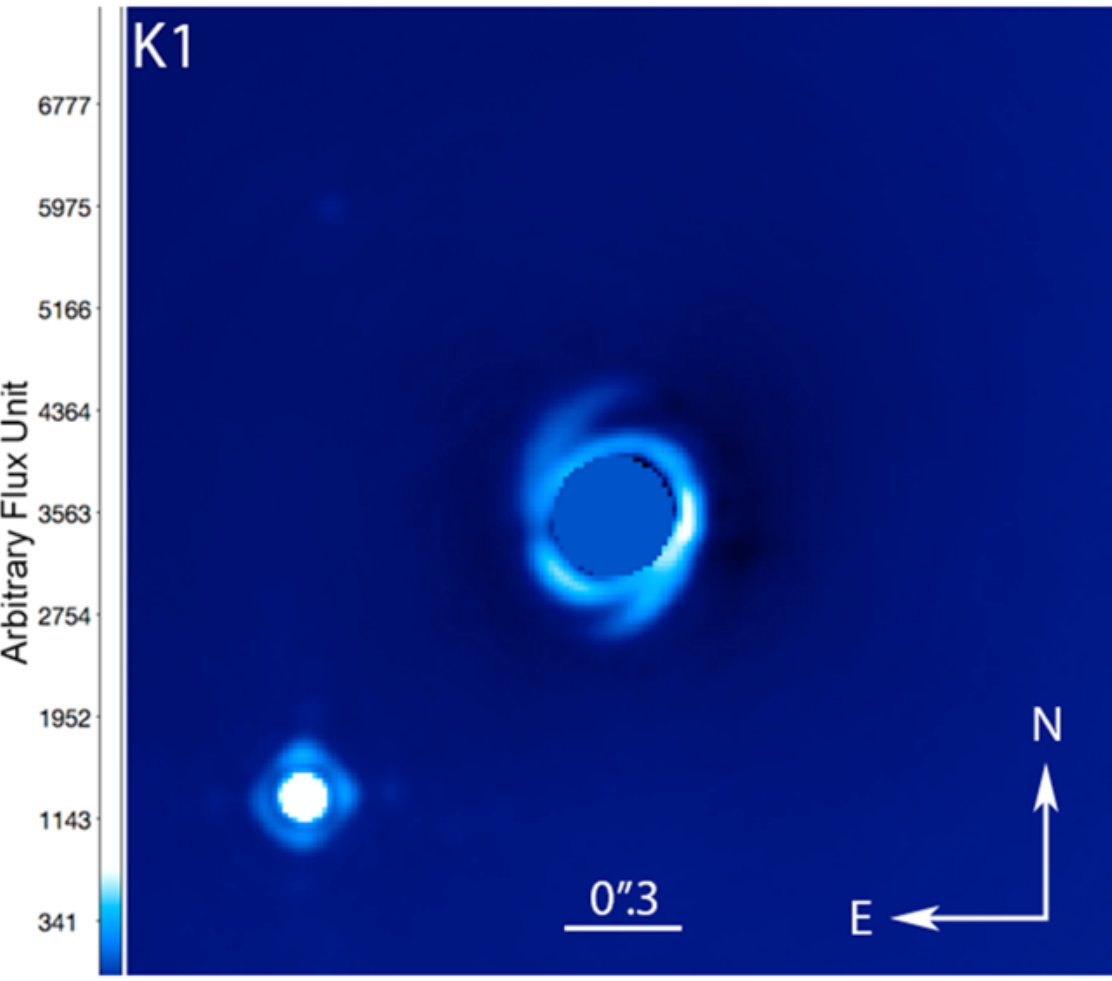}{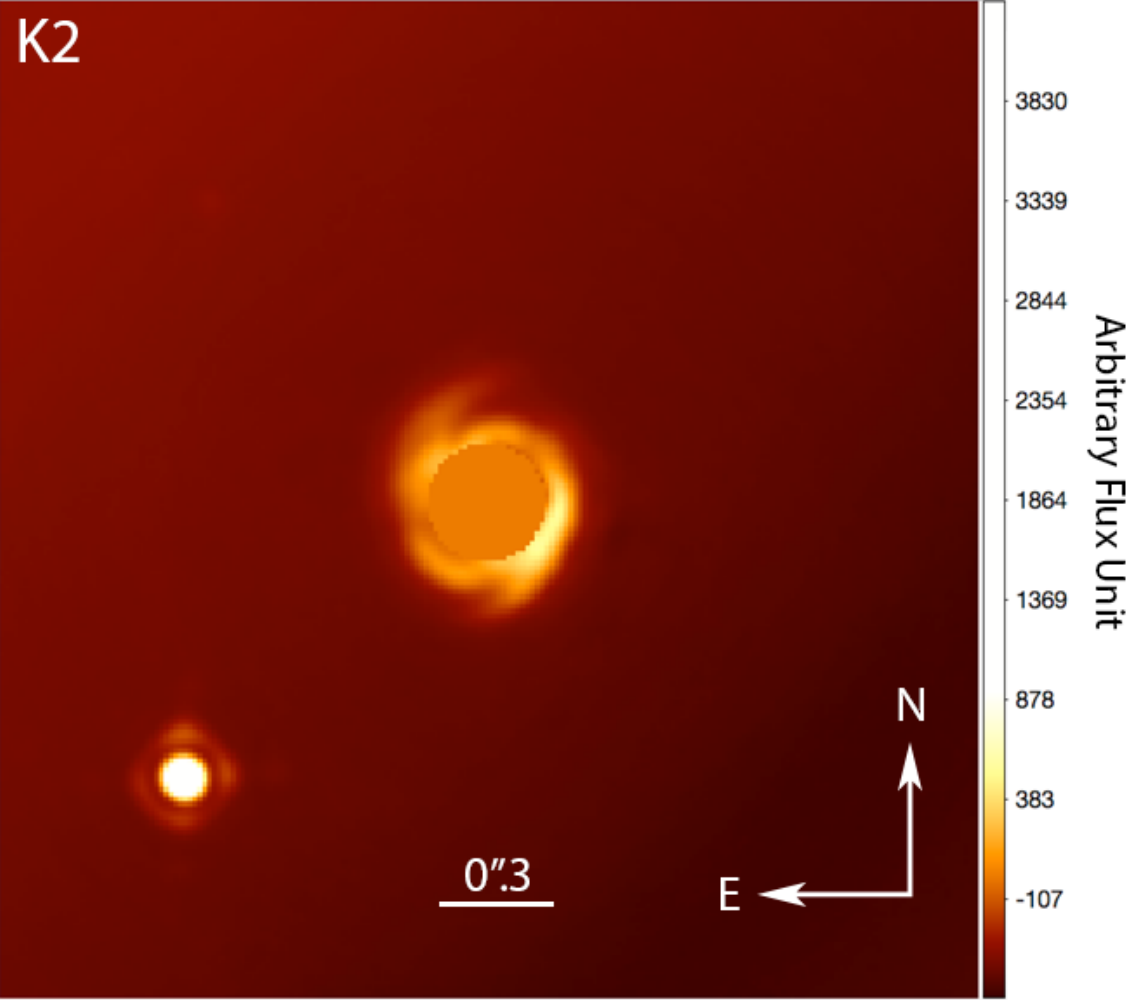}
\caption{\footnotesize PSF subtracted IRDIS exposure in K1 (left) and K2 (right) of the HD 100453 system showing the spiral disk, the companion M dwarf at separation 1$\farcs$05, PA=132$\pm 1^{o}$, as well as background star C (faintly) at 1$\farcs$05, PA=43$\pm 1^{o}$. \label{spiral}}
\end{figure}

\begin{figure}[htpb]
\figurenum{2}
\epsscale{1}
\plotone{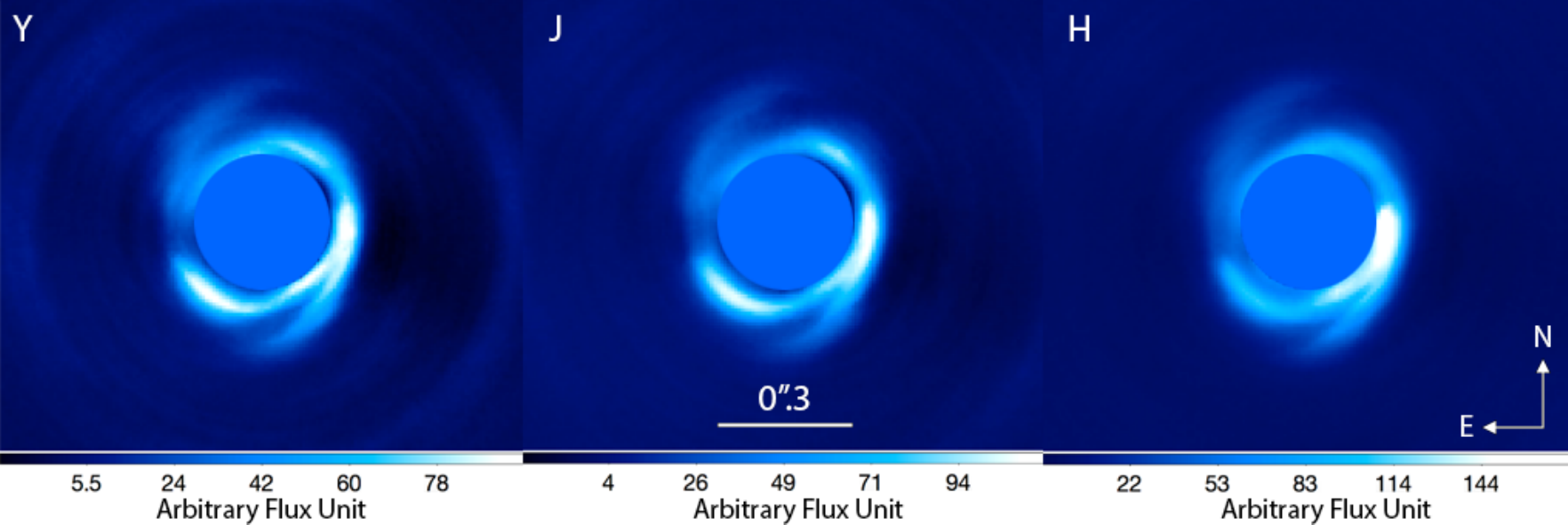}
\caption{\footnotesize PSF-subtracted Y$-$, J$-$, and H-band IFS images of HD 100453, verifying the detection of the spiral disk in each band.  \label{IFS}}
\end{figure}

We also reduced the IRDIS data through analysis and subtraction of the principal components of the PSF via the KLIP method (described in detail by \citealt{soummer12}) using self-developed IDL routines (\cite{hansonapai2015}, Apai et al., submitted). This method relies on the rotation of the field to compare radial segments of each target frame to those taken with at least of  1.5$\times$FWHM pixels separation to achieve the best subtraction of principal components. Over the course of our observations the field rotated by 12.5$^{o}$. The KLIP method is optimized at finding point sources, and does well to detect HD 100453 B and another background star, but partially subtracts the nearly face-on disk. No other objects were detected in this reduction, which we discuss in $\S$4.2.

\section{Results}

\subsection{The spiral disk}

Our primary result is the detection of a ring surrounded by two spiral arms that is clearly resolved in Y$-$, J$-$, H$-$, and K-band scattered light, shown in Figures 1 \& 2. Our observations trace the disk to 0$\farcs$37 ($\sim$42 AU) from the star with  fractional luminosity ($f=L_{disk}/L_{\star}$) of $f_{Y}=0.0127, f_{J}=0.0209, f_{H}=0.068, f_{K1}=0.049,$ and $f_{K2}=0.103$\color{black}. The apparent furthest extent of the ring is $\sim$0$\farcs$25 ($\sim$29 AU) in the NW and SE. Assuming a circular geometry, we calculate the inclination of the ring to be $\sim$34$^{o}$ from face on. The NE spiral intersects the ring between PA=22$^{o}$ and PA=79$^{o}$, while the arm extends to PA=$-$15$^{o}$. The SW spiral intersects the inner ring between PA=194$^{o}$ and PA=251$^{o}$, while the arm extends to PA=155$^{o}$. The apparent opening angles of the NE and SW spiral arms are $\sim 38^{o}$ and $\sim 30^{o}$, respectively.

\begin{figure}[htpb]
\figurenum{3}
\epsscale{1}
\plotone{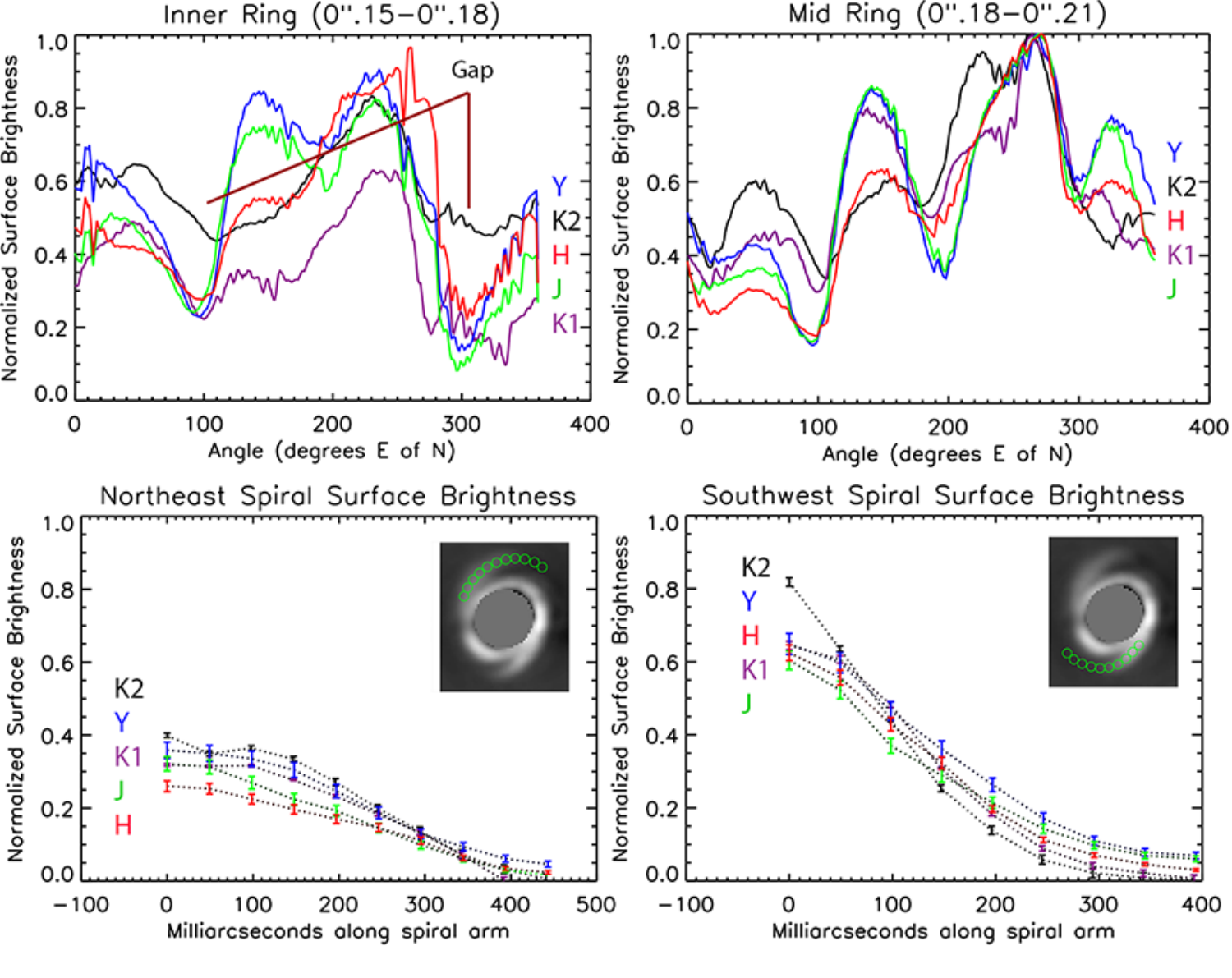}
\caption{\footnotesize (Top row:) Azimuthal brightness of the ring, taken at half degree increments and smoothed by a running boxcar. (Bottom row:) Surface rightness profile in the spiral arms, with the K1 IRDIS image overlayed to show the apertures used to extract the surface brightness. Each image is normalized independently.\label{ring}}
\end{figure}

The 0$\farcs$18 - 0$\farcs$25 ($\sim$21-29 AU) ring displays several finer structures, illustrated in the azimuthal surface brightness profiles in Figure 3.  First, we identify sharp axisymmetric drops in brightness to the Northwest and Southeast, extending $\sim$0$\farcs$03 outside of the inner working angle for more than 30$^{o}$ on each side, which we interpret as the signature of a gap in the disk. It is important to note that our inner working angle of 150 mas is outside of the actual coronagraphic mask, whose radius is about 90 mas. The edge of the mask is affected by scattering and residual light, and hence the inner working angle is loosely defined as the radius at which significant contrast may be obtained. Still, the dark crescents that we observe are what would be expected of a gapped disk with a wall at $\sim$21 AU projected separation at an inclination of $\sim$34$^{o}$.

We identify several asymmetric features in the disk. In particular, we identify a sharp drop in intensity at PA$\sim$100$^{o}$ in all wavelengths, and two peaks in intensity at PA$\sim$140$^{o}$ and PA$\sim$250$^{o}$. The peak at PA$\sim$140$^{o}$ appears to be highly color-dependent. The SW spiral arm also appears to be $\sim$2$\times$ brighter and $\sim$25\% less extended than the NE arm. These features are likely the combined result of the disk inclination (i.e. scattering angle), differences in grain populations, differential illumination by the M dwarf companion, and/or scale height variations induced by planets within the gap, similar to those seen induced by the planet in $\beta$ Pictoris \citep{apai15}. The exact combination and degeneracy between these parameters remains to be explored through detailed radiative transfer modelling. 
\color{black}
\section{Discussion}

\subsection{Disk Morphology}

We detect a multitude of features which may help to infer the nature of an undetected planetary system in HD 100453 $-$ i.e. the spiral arms, the non-axisymmetric bright regions of the ring, the null-detection of giant planets at wide separation, and the possible gap. The spiral disk is similar in size to the orbits of Neptune and Pluto in our own Solar system, and hence it may be relevant as an example of early Kuiper belt-type structures. Compared to other two-armed spiral disks (e.g. SAO 206462, \citealt{garufi13}, MWC 758; \citealt{benisty15}) the one in HD 100453 appears to be highly symmetric, with the two spiral arms departing the ring approximately 180$^{o}$ apart. We present a side-by-side comparison of these three well-resolved disks in Figure 4. Together, these constitute the best sample of directly imaged two-armed spiral protoplanetary disks. The addition of HD 100453 to the sample may provide interesting insights to our understanding of the evolution of two-armed spirals. In particular, the spiral arms in HD 100453 are highly symmetric in their azimuthal extension around the star compared to SAO 206462 and MWC 758 and exhibit greater opening angles than those in SAO 206462, MWC 758, or those predicted by any models. These features provide important comparisons for theoretical modelling studies. 

\begin{figure}[htpb]
\figurenum{4}
\epsscale{1}
\plotone{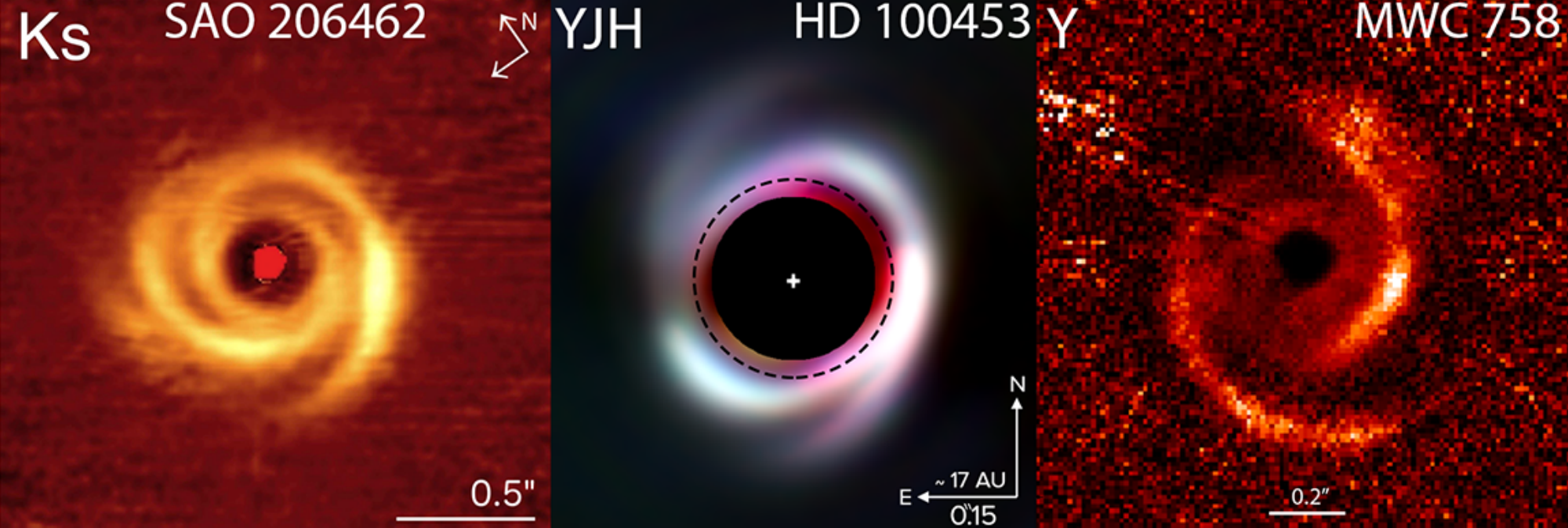}
\caption{\footnotesize  Comparison of the three spiral disks around young stars resolved in high contrast imagery. The spiral disk in HD 100453 is the smallest of these three, both in angular and projected size, and displays a high degree of symmetry in the arms as well as several asymmetric bright features in the ring. Presented here are the images of SAO 206462 \citep{garufi13}, and MWC 758 \citep{benisty15} alongside our tri-color (RGB=HJY) image of HD 100453. The images of HD 100453 in the three bands have been stretched independently to display a similar contrast. The black mask in the center hides the regions that are significantly affected by the coronagraph, while the dashed circle shows the innermost radius at which structures are considered to be real (0$\farcs$15). The SAO 206462 and MWC 758 images are reprinted with permission of their respective authors and the editor of \textit{Astronomy \& Astrophysics.}}
\end{figure}

\cite{muto12}, \cite{grady13}, and \cite{benisty15} reproduce the spiral structures in SAO 206462 and MWC 758 through spiral wave density theory, with two planetary perturbers inside of the spiral arms inciting the two spiral structures. In contrast to the spiral density wave theory models, \cite{juhasz15} find through detailed radiative transfer and hydrodynamic simulations that spirals launched as density perturbations require a relative change of about 3.5 in the surface density to be detectable using current telescopes and instrumentation $-$ a factor of $\sim$8 times higher than seen in hydrodynamic simulations. Rather, \cite{juhasz15} also find that spirals launched as a change of 20\% in pressure scale height should indeed be detectable, and suggest that presently observed spirals in protoplanetary disks are the result of changes to the vertical structure of the disk. Interestingly, \cite{pohl15} find that highly symmetric two-armed spirals may only be reproduced with two planets with very specific mass ratios and locations of the planets inside of the arms. As predicted for other spirals, an intriguing possibility for HD 100453 is that unseen planets exist inside of our inner working angle, which may be perturbing the disk from the inside.

The fact that this area of the disk is well beyond the snow line gives rise to significant potential for giant planet formation, which may have opened one or more gaps in the inner regions of the disk, like those seen in other disks (e.g. HD 169142, \citealt{momose15}; SAO 206462, \citealt{garufi13}; V 1247 Ori, \citealt{kraus13}, etc.). Although the coronagraphic mask limits our ability to observe inwards of $\sim$17 AU from the star, there is indeed evidence of a gap extending to $\sim$21 AU projected separation in our observations, which corresponds to the predicted location of the gap wall at 17$\pm$2 AU found by SED modelling in \cite{maaskant14}. Additionally, the brightness of the $\sim$21-29 AU ring may be further evidence of an inner gap in the disk structure, as features at such large separation are likely to be in shadow of the inner regions of the disk. However, the near infrared flux of the object (see \citealt{maaskant14}) suggests the presence of an inner disk extending to the dust sublimation radius that is unresolved in our observations. The NIR excess, along with the brightness of the outer disk ring, is suggestive of a close inner gap wall ($\lesssim$1 AU), and a gap size comparable to suggestions for the inner gap in HD 169142 (\citealt{wagner15} and \citealt{osorio14}). 

\subsection{Companions and other objects in the field of view}

Our observations show another object at 1$\farcs$05, PA=43.1$\pm 1^{o}$, and contrast relative to HD 100453 A of $\Delta$K1=10.51$\pm$0.04 mag and $\Delta$K2=10.57$\pm$0.05 mag. The object appears in the 2003 HST ACS/HRC images at $\sim $0$\farcs$8, PA=30$^{o}$ and is characterized as a background object in \cite{collins09} from a 3 month baseline with VLT/NACO. Our observations extend the baseline to 12 years, and confidently rule out star C as a candidate co-moving member of the HD 100453 system. In June 2003 \cite{chen06} detect HD 100453 B at 1$\farcs$05, PA=127$\pm 1^{o}$. In our April 2015 observations we detect the companion at 1$\farcs$05 and PA=132$\pm 1^{o}$, giving 5$\pm 2^{o}$ orbital motion over 12 years. Assuming an approximately face on orbit, we estimate a period of $\sim$850 yr, comparable to the estimates of $\sim$930 yr \citep{chen06} under the assumption of a 0.3 M$_{\sun}$ companion on a circular orbit. In Figure 5 we show the relative positions of HD 100453 and star C in 2003 and 2015 overlayed on the IRDIS K1 image.

\begin{figure}[htpb]
\figurenum{5}
\epsscale{.4}
\plotone{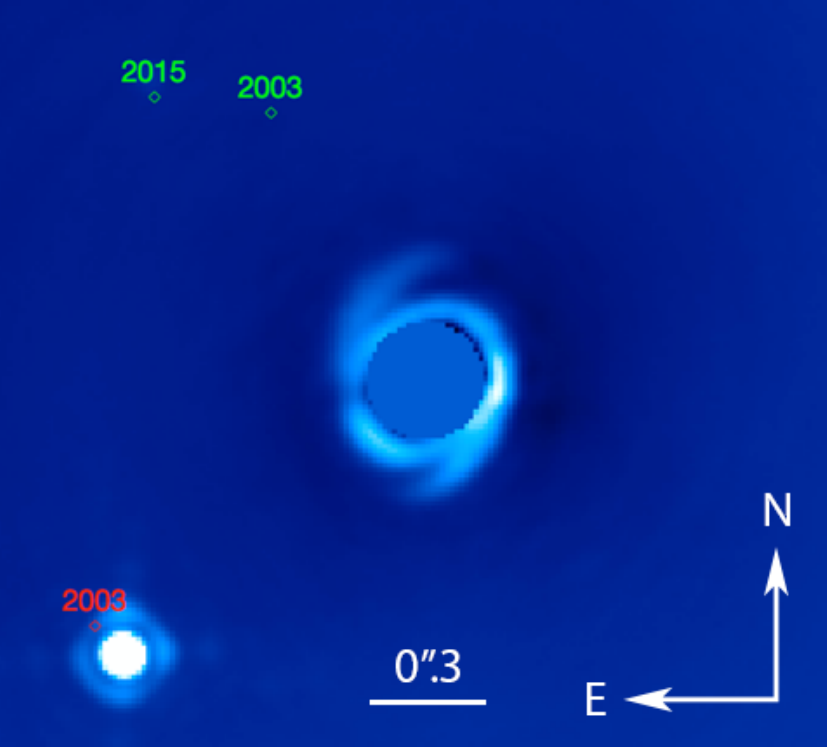}
\caption{\footnotesize IRDIS K1 image of the HD 100453 system, showing the location of star C in 2003 and 2015 labeled in green at image top, and the location of HD 100453 B in red, $\sim$5$^{o}$ from where it appears in our image.}
\end{figure}

Apart from HD 100453 B and the background star “C” we do not see evidence for a companion candidate in the field of view centered on HD 100453. The observations presented here are part of a larger survey to assess the frequency of large-separation giant planets and will be included in an upcoming, detailed study to accurately characterize the survey's sensitivity and the constraints it places on giant planet occurrence in A-stars. Nevertheless, as the potential presence of giant planets in the HD 100453 disk may be relevant for interpreting the two-armed spiral disk presented in this study we provide a preliminary estimate for our sensitivity. 

We injected artificial planets (point spread functions) of varying intensity and at various locations to the KLIP-reduced images to identify the approximate brightness levels as a function of separation where candidate planets are detectable. Our tests suggest that we can detect, as candidates, planets with a $\Delta$K1 contrast of 14.3 mag at a stellocentric separation of 0$\farcs$5, corresponding approximately to the outer edge of the spirals. Assuming a distance of $\sim$110 pc and an age of 10 and 20 Myr, we can compare our sensitivity limits to the K-band brightness of planets predicted by the hot start models of \cite{baraffe98}. Our comparison suggests that our sensitivity is sufficient to detect 3 M$_{Jup}$ (for an age of 10~Myr) or 4 M$_{Jup}$ (for an age of 20~Myr) at 0$\farcs$5. Although this analysis is preliminary, it argues against the presence of a super-Jupiter comparable to HR 8799bcde or $\beta$~Pictoris~b, outside the two-armed spiral disk imaged around HD 100453. 

\section{Conclusions}

In this letter we've presented the first detection of the circumstellar disk in HD 100453 in high contrast imagery. Our observations trace the disk to 0$\farcs$37 ($\sim$42 AU) from the star with fractional luminosity of $f_{Y}=0.0127, f_{J}=0.0209, f_{H}=0.068, f_{K1}=0.049,$ and $f_{K2}=0.103$. The images show a ring of dust at $\sim$21-29 AU projected separation from the star, with two highly symmetric spiral arms to the Southwest and Northeast, which we estimate to be inclined $\sim$34$^{o}$ from face on by assuming a stellocentric symmetry of the ring. Our observations also show evidence for a gap extending to 0$\farcs$18 ($\sim$21 AU projected separation), seen as dark crescents in the NW and SE, as well as several asymmetric brightness features in the disk. \color{black} We observe $\sim$5$^{o}$ orbital motion of HD 100453 B between 2003-2015, and estimate a period of $\sim$850 yr. This discovery of the spiral disk in HD 100453 adds to the current small sample of well-resolved two-armed spiral circumstellar disks, and contributes important details that may aid in our understanding of their origin and evolution. 

\section{Acknowledgments}

This work is based on observations performed with VLT/SPHERE under program ID 095.C-0389A and supported by the National Science Foundation Graduate Research Fellowship Program under Grant No. 2015209499. The results reported herein benefited from collaborations and/or information exchange within NASA's Nexus for Exoplanet System Science (NExSS) research coordination network sponsored by NASA's Science Mission Directorate. We would also like to thank the astronomers at the VLT who carried out our observations in service mode.

\end{document}